\documentclass[12pt]{article}
\usepackage[latin1]{inputenc} 
\usepackage{graphicx}
\usepackage{sidecap}
\usepackage{wrapfig} 
\usepackage{subfig}
\usepackage{float}
\usepackage{amsmath} 
\usepackage{amsfonts} 
\usepackage[mathscr]{eucal}
\begin{document} 
\title{${\rm SU}(N) \to {\rm Z}(N)$ dual superconductor models: the magnetic loop ensemble point of view} 
\author{Luis E. Oxman \\ \\
Instituto de F\'{\i}sica, Universidade Federal Fluminense,\\  
Campus da Praia Vermelha, Niter\'oi, 24210-340, RJ, Brazil. }

\maketitle

\begin{abstract}

In this work, we initially discuss some physical properties of effective ${\rm SU}(N) \to {\rm Z}(N)$ YMH models,  emphasizing the important role of valence gluons. Next, we review how adjoint fields are naturally generated as an effective description of ``adjoint'' loops  in $4D$. Finally, we 
discuss the consequences that can be learnt from this point of view, and  briefly comment on some improvements. 

\end{abstract}

\section{Introduction} 
\label{intro}

 Dual superconductivity is a promising scenario to understand confinement in pure Yang-Mills (YM) theories \cite{N}-\cite{3}. On the one hand, many different dual superconductor models where the confining string is represented by a {\it classical} vortex solution have been explored (see \cite{Su-90}-\cite{conf-qg} and references therein). On the other, results obtained from Monte Carlo simulations have been understood in terms of the proliferation of magnetic degrees of freedom, detected in the lattice (see \cite{ref19}  \cite{book-G} and references therein). The latter correspond to ensembles of {\it quantum} objects such as center vortices, monopoles and chains, which capture the path-integral measure of pure YM.  
These approaches can be thought of as (dual non-Abelian) versions of the understanding of superconductors in terms of the Guinzburg-Landau wave functional and the condensation of Cooper pairs, respectively.
Analyzing to what extent the existing lattice phenomenology is accommodated and the relation with well-established {\it quantum} ensemble descriptions are two important tasks to be pursued. 

On the side of effective models, those based in the bosonic sector of supersymmetric theories have been extensively studied. In particular,
non-Abelian strings in ${\rm U}(N)_{\rm gauge}  \times {\rm SU}(N)_{\rm flavor}$ YMH models with fundamental Higgs fields were introduced in refs. \cite{David}, \cite{it},  \cite{GSY}. In the SSB phase, there is a remnant global flavor-locking ${\rm SU}(N)_{\rm C + F}$ symmetry which equips  confining strings with non Abelian degrees of freedom. 
In ref. \cite{GSY}, a non-supersymmetric model with ${\rm SU}(N)\times {\rm U}(1)$ gauge symmetry and a similar color-flavor locking vacuum was introduced, with decay rates of quasi-stable $k$-strings \cite{AS}  similar to those present in large $N$ pure YM. Although it is constructed in terms of $N$ flavors of fundamental Higgs fields, the presence of the ${\rm U}(1)$ factor leads to good $N$-ality properties. 
 As is well-known, this important property (see ref. \cite{Biagio}) can also be implemented in models with at least $N$ flavors of adjoint Higgs fields \cite{deVega}-\cite{HV}. In ref. \cite{Marco}, they were  
implemented in the bosonic part of  ${\cal N}=2$ super QCD with a soft breaking mass term. 
Relying on phenomenological arguments, they were  proposed to describe the confining string in pure YM (see refs. \cite{Su-90}-\cite{Baker-Dosch} and references therein). In ref. \cite{conf-qg}, we pointed out that ensembles of monopoles carrying adjoint charges could be naturally represented by models based on a set of adjoint Higgs fields. This type of magnetic degree has been detected in lattice calculations of pure YM, interpolating different center vortices to form closed chains \cite{book-G}. In ref. \cite{OST}, assuming phenomenological information to characterize an ensemble of magnetic ``adjoint'' loops, such as tension, stiffness  and (magnetic) color degrees of freedom (d.o.f.), we showed how adjoint fields can be generated. 

In this talk, we shall  briefly review some of these ideas and 
discuss the consequences that can be learnt from this point of view. We shall also comment on some improvements in the description of non Abelian ensembles of loops; a  detailed discussion will be reported elsewhere.

\section{$ {\rm SU}(N) \to {\rm Z}(N)$ models with adjoint scalars}
\label{ZNmod}

Let us consider a general Yang-Mills-Higgs (YMH) model with a set of adjoint fields $\psi_I \in \mathfrak{su}(N)$,
\begin{eqnarray}
&&\frac{1}{2} \langle D_\mu \psi_I , D^\mu \psi_I\rangle +
\frac{1}{4g^2} \langle F_{\mu \nu}-J_{\mu \nu}, F^{\mu \nu}-J^{\mu \nu}\rangle - V_{\rm Higgs}(\psi_I) \;,  
\label{gen-mod}
\\
&& ~~~~~~~  D_\mu=\partial_\mu  + \Lambda_\mu  \wedge \makebox[.5in]{,}
F_{\mu \nu}= \partial_\mu \Lambda_\nu -\partial_\nu \Lambda_\mu + \Lambda_\mu \wedge   \Lambda_\nu \;, \nonumber
\end{eqnarray}
where $\Lambda_\mu $ is a non Abelian (dual) gauge field and $I$ is a flavor index; the color index can be made explicit by expanding the fields in a Lie algebra basis $T_A$, $\psi_I = \psi_I^A \, T_A$.  We are using the notation $X \wedge Y = -i\, [X, Y]$, and $\langle X, Y \rangle$ for the Killing form.
The source, $J_{\mu \nu} = \vec{J}_{\mu \nu}|_q\, T_q$~, ~ $ \vec{J}_{\mu \nu}= 2\pi \, 2N\, \vec{w}_{\rm e}\, s_{\mu \nu}$ depends on a weight $\vec{w}_{\rm e}$ of the quark representation.\footnote{A weight $\vec{w}$ is defined by the eigenvalues of the Cartan generators $T_q$ corresponding to one common eigenvector.  
\[
[T_q,T_p]=0  \makebox[.5in]{,} T_q\, eigenvector = \vec{w}|_q\,  eigenvector  \makebox[.5in]{,} q,p= 1, \dots, N-1\;.
\]
}
Each static source contributes to $s_{\mu \nu}$ with, 
\[
s_{0i} = 0  \makebox[.5in]{,} s_{ij} = - \epsilon_{ijk}\, \int ds\, \frac{dx_k}{ds}\,  \delta^{(3)}(x-x(s)) \;,
\] 
where $x(\sigma)$ is a Dirac line ending (starting) at the quark location. Higgs potentials that drive ${\rm SU}(N) \to {\rm Z}(N)$ are characterized by configurations of absolute minima with the property $U \psi_I^0 U^{-1} = \psi_I^0$ just if $U \in {\rm Z}(N)$, that is, ${\cal M}={\rm SU}(N)/{\rm Z}(N)$. As is well known, this phase is characterized by: 

\vspace{.2cm} 

$\bullet$ {\bf $N$-ality}: This is a consequence of $\Pi_1 ({\cal M}) = {\rm Z}(N)$.  The asymptotic behavior of a center string is {\it locally} a pure gauge (but not {\it globally}) that can be written in terms of the phase,
\begin{equation}
S= e^{i\varphi\, \vec{\beta} \cdot \vec{T}}   \makebox[.5in]{,} \vec{\beta} = 2N\, \vec{w} \makebox[.5in]{,} \vec{\beta}\cdot \vec{T} = \vec{\beta}|_q T_q \;.
\label{beta}
\end{equation}
For the fundamental representation, there are $N$ weights $\vec{w}_i$ (fundamental colors). 
Infinite adjoint strings are trivial, as the asymptotic behavior $S\sim  e^{i\varphi\, 2N\, \vec{\alpha} \cdot \vec{T}}$ ($\vec{\alpha}$ a root)  is a closed loop in ${\rm SU}(N)$. Then, it  can be continuously deformed to avoid any defect at the origin ($\Pi_1({\rm SU}(N))=0$).\footnote{The roots are the weights of the adjoint representation, which acts via commutators 
$
[T_q,E_\alpha]= \vec{\alpha}|_q\, E_\alpha $.}   

 \vspace{.2cm}

\begin{figure}[h] 
\hspace{0.1cm}
\centering  
(a) \hspace{-1cm}
\includegraphics[scale=.11, bb = -290 0 400 850]{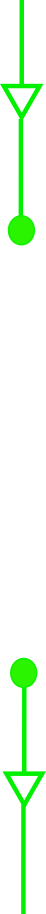}
\hspace{1cm}
(b) \hspace{-1.4cm}
\includegraphics[scale=.11, bb = -290 0 300 909]{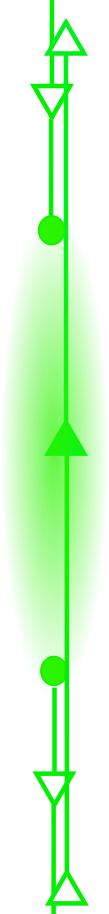} 
\hspace{1cm} 
(c) \hspace{-1.4cm}
\includegraphics[scale=.12, bb = -370 -120 600 750]{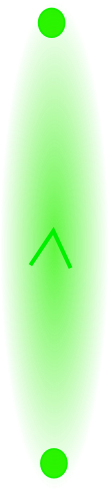} 
\caption{Here, we show: (a) the external Dirac lines, (b) their cancellation against similar terms in $F_{ij}$ and (c) the final induced finite string.}
\label{fig-1-en}       
\end{figure}

$\bullet$ {\bf Normal mesons}: A pair of quark sources, with fundamental weights $\vec{w}, -\vec{w} $, are introduced by the Dirac lines in fig. \ref{fig-1-en}a.  The absolute minimum is characterized by $S$ in eq. (\ref{beta}).  Around the Dirac lines, the field profiles should be close to the true vacuum. In this way, the field configuration will be a (singular) pure gauge in that region, and the string-like singularity in $F_{ij}$ cancels the Dirac lines (see fig. \ref{fig-1-en}b). On the other hand, between the quarks, the phase defect implies a false vacuum region. At the end, a finite center string with smooth energy density  is induced 
 (see fig. \ref{fig-1-en}c).  

\subsection{Additional physical properties} 

\vspace{.2cm}

$\bullet$ {\bf Hybrid mesons}: As shown in refs. \cite{HDavid}, \cite{Dav}, non-Abelian string models have solutions where different strings can be interpolated by monopoles (forming complexes). 
The latter were interpreted as confined string-attached gluons \cite{GSY}, \cite{Nonajun} (for a similar interpretation in the context of supersymmetric theories, see ref. \cite{Mgluon}). These configurations were discussed for the first time in ref. \cite{HK}, in the ${\rm SU}(2)$ case.  
In ref. \cite{conf-qg}, motivated by previous work on the description of chains in the continuum, using defects of a local Lie basis \cite{L2008}, an analysis of center strings interpolated by monopoles was presented, pointing to some physical consequences. In this regard, lattice calculations predict a rich spectrum of exotic mesons. Some of them correspond to  $qg\bar{q}'$ hybrids, where a nonsinglet color pair and a valence gluon form a colorless state.  This state is currently searched by a collaboration based at the Jefferson Lab (GlueX) \cite{gluex}.  

The (infinite)  hybrid glue  is induced by fundamental sources $\vec{w}, -\vec{w}' $, with $\vec{w} \neq \vec{w}'$. This state is associated with a non Abelian phase ($\theta$, $\phi$ are spherical angles),
\[
S= e^{i\varphi\, \vec{\beta} \cdot \vec{T}}\, W(\theta)
\makebox[.5in]{,}
W(\theta) = e^{i\theta \,\sqrt{N} T_{\alpha} } \makebox[.5in]{,}  \vec{\alpha}=  \vec{w} -\vec{w}' \;.
\] 
Around the north pole, 
$
S \sim  e^{i\varphi\, \vec{\beta} \cdot \vec{T}} 
$,
while around the south pole, using that $W(\pi)$ is a Weyl reflection,
we get,
$
S \sim  W(\pi) \,e^{i\varphi\, \vec{\beta}' \cdot \vec{T}} 
$. As gauge transformations act on the left, this is locally equivalent to the behavior $\sim e^{i\varphi\, \vec{\beta}' \cdot \vec{T}}$. Note that $S$ gives a well-defined mapping for the local Cartan directions on $S^ 2$, 
$ n_q=ST_q S^ {-1} $, with a point-like defect at the origin. Of course, in the complete ansatz, some profile functions must tend to zero at this singularity, and others must tend to zero on those line-like singularities where the guiding centers of the strings are located. The gauge invariant monopole charge is obtained from the (dual) field strength projection along the {\it local} Cartan directions $n_q$,
$
\vec{Q}_m=   2\pi\, 2N\, (\vec{w} -\vec{w}') =  2\pi\, 2N\, \vec{\alpha}\;. 
$
This is a root, or weight of the adjoint representation, thus leading the dual monopole to be identified with a  valence gluon with adjoint color $\vec{\alpha}=\vec{w} -\vec{w}'$.

\vspace{.2cm}

$\bullet$ {\bf Valence gluons are confined}: Another important property of the {\it dynamical} dual monopole is that it cannot exist as an isolated object. This comes about as ${\cal M} = {\rm Ad}({\rm SU}(N)) $ is a compact group, so that $\Pi_2({\cal M})= 0$. 
 
\vspace{.2cm}

\begin{figure}[h] 
\hspace{0.1cm}
\centering  
(a) \hspace{-1cm}
\includegraphics[scale=.13, bb = -290 0 400 850]{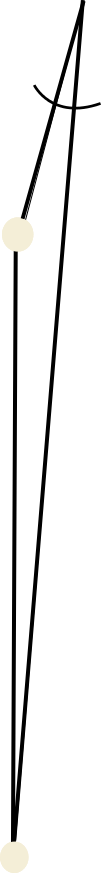}
\hspace{1cm}
(b) \hspace{-1cm}
\includegraphics[scale=.13, bb = -290 0 300 909]{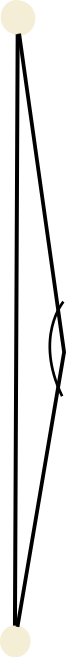} 
\hspace{1cm} 
(c) \hspace{-1cm}
\includegraphics[scale=.11, bb = -370 0 600 750]{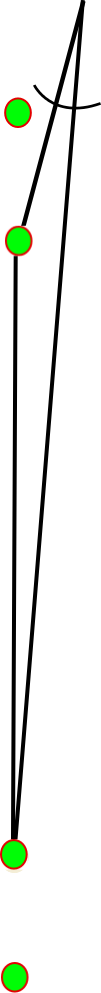} 
\caption{Bipolar coordinates with foci at: a pair of adjoint quarks, (a) $\gamma \sim 0$,  (b) $\gamma \sim \pi$,  and at inner quarks in double pairs of fundamental quarks (c).}
\label{fig-a-en}       
\end{figure}

$\bullet$ {\bf Adjoint quarks are not confined}:  Consider a pair of adjoint sources at a finite distance, with  weights  $\vec{\alpha}, -\vec{\alpha}$, respectively. We could proceed as for normal mesons, considering
$
S_{\rm Abe}= e^{i\varphi\, 2N\, \vec{\alpha} \cdot \vec{T}} $, $ \vec{\alpha}=\vec{w} -\vec{w}' 
$, 
thus obtaining a linearly growing (finite) energy. However, at some point, another type of configuration will be preferred. We can also consider the non Abelian phase  \cite{DL},
$S_{\rm non-Abe} =  e^{i\varphi\, \vec{\beta} \cdot \vec{T}}  W(\gamma) \,   e^{-i\varphi\, \vec{\beta}'  \cdot \vec{T}}$, 
where $\gamma$, $\varphi$ are bipolar angles. Around $\gamma \sim 0$ (see fig. \ref{fig-a-en}a),  
$S_{\rm non-Abe} \sim e^{i\varphi\, 2N\, \vec{\alpha} \cdot \vec{T}}$. Then, to cancel the Dirac lines that run between the sources and infinity, the field profiles must tend to a true vacuum in that region. 
%
%
In addition, for $\gamma \sim \pi $ 
(see fig. \ref{fig-a-en}b), $S_{\rm non-Abe} \sim  W(\pi) $. That is, between the quarks there is no phase defect (nor Dirac lines) and the fields can also assume values close to the true vacuum. Then, the energy density will only be significant around the quarks, where  $S_{\rm non-Abe}$ looks like the non Abelian phase that characterizes a valence gluon (in spherical coordinates around each one of the sources). 
In other words, the energy minimization will induce a dynamical dual monopole around each adjoint quark, screening them by the formation of a pair of adjoint-quark/valence-gluon bound states. This is the correct picture for the breaking of the adjoint string \cite{dFK}.

\vspace{.2cm}

$\bullet$ {\bf Difference-in-areas for doubled pairs of fundamental quarks}: 
%
%
%
In ref. \cite{GR}, it was shown that double-winding Wilson loops in SU(2) gauge theory, computed by Monte Carlo simulations, obey a  difference-in-areas law, in agreement with the center vortex model of confinement.
On the other hand, Abelian pictures lead to 
a sum of areas (see \cite{GR} and references therein).  Here, we will show that ${\rm SU}(2) \to {\rm Z}(2)$ effective YMH models also agree with Monte Carlo simulations. For this aim, let us consider four fundamental ${\rm SU}(2)$ quarks as shown in fig. \ref{fig-a-en}c. The two upper (lower) quarks have negative (positive) weight\footnote{For  {\rm SU}(2), weights are one-component. The fundamental ones are $\pm w$, $w= \frac{1}{2\sqrt{2}}.$}, doubling the simple scheme in fig. \ref{fig-1-en}a. 
When the upper and lower sets are sufficiently apart,  the absolute minima will be given by  
$
 S_{\rm non-Abe} = e^{i\varphi\, \beta T_1}\,  W(\gamma) \;  e^{i\varphi\, \beta  T_1}  $. 
When $\gamma \sim 0$, the behavior 
$S_{\rm non-Abe} \sim e^{i\varphi\, 2\beta T_1}$ will introduce a doubled string-like singularity in $F_{ij}$. This will cancel the doubled Dirac lines between the outer quarks and infinity (region I), and will partially cancel the single Dirac line between the upper (lower) quarks (region II). .When  
$\gamma \sim \pi $ we have the regular behavior  $S_{\rm non-Abe} \sim  W(\pi) ~$, and no Dirac lines (region III). Then, the smooth energy density requirement leads to true vacuum in regions I and III. In addition,  there will be a state similar to a fundamental string between the upper (lower) quarks  with lengths $L_{\rm u}$ ($L_{\rm l}$). In other words, the generated action for this static configurations goes like
$ T L_{\rm u}+T L_{\rm l}  = T L_{o} - T L_{i}$, where $L_{\rm o}$  ($L_{\rm i}$) is the distance between the outer (inner) quarks, and  $T$ is the (infinite) configuration lifetime.

The possibility of a fundamental  string between, say, the upper quarks (each carrying positive charge $w$) is due to the induction of a valence gluon (with adjoint charge $-2w$) around the inner quark.  This partial screening also occurs at the inner lower quark location.

\section{Ensembles and fields}  
\label{ensembles}
 
Lattice calculations in pure YM show that  ensembles of magnetic  objects capture the main contribution to the path-integral in the infrared regime (see \cite{greensite,book-G,ref11} and references therein). In particular, the center vortices that arise in lattice center gauges are good at describing $N$-ality and have a physical density scaling.  A large percentage of them end at monopoles forming closed chains (see \cite{book-G} and references therein). Similarly to the dual description of valence gluons,  the monopole component in chains can be described by Weyl transformations, so they carry adjoint charges labelled by the roots of $\mathfrak{su}(N)$. Here, we review 
the description of ensembles of  loops that carry an adjoint charge, equipped with non Abelian (magnetic) color d.o.f. \cite{OST}. We also clarify some points regarding the projection over a reduced sector of well-defined color states. 

As is well-known, ensembles of one-dimensional objects lead in general to effective field models \cite{bar-sam}-\cite{HSi}. The sum over loops can
be seen as a sum over different numbers of particle worldlines, that is, a second quantized field theory represented by a path-integral.   
The loops can  be characterized by  phenomenological properties: tension $\tau $, stiffness $\xi^{-1}$ (which is important for the continuum limit),  interactions among them, etc. We also considered the coupling,
\begin{equation}
\int ds\, \left[\frac{1}{2} (\bar{z}_c
\dot{z}_c - \dot{\bar{z}}_c z_c)    - i\,  u_\mu\,  I_{A}\, \Lambda_{\mu}^{A}(x(s))\right]
\makebox[.3in],I_A = M_A |_{cd} \,
\bar{z}_{c} z_{d} 
\label{Balac}
\end{equation}
which was introduced in ref. \cite{ref26} to describe a classical relativistic  particle  interacting with a non Abelian gauge field. In our work, it was included to make contact with typical terms in the dual effective YMH models. The $z_a$'s, $a=1, \dots,   D $, are complex variables; $D$ is the dimension of the group representation under consideration. In particular, for adjoint loops we have $D=N^2-1$ and the matrix elements of  $M_A = R(T_A)$, $A=1,\dots, N^2-1$, are proportional to the $\mathfrak{su}(N)$ structure constants.   The partition function for the ensemble of adjoint loops was then defined by, 
\begin{eqnarray} 
&& ~~~~~~~~~~~~~~~~~~~~~~~~~~ Z=\int [D\phi] [D\phi_A] \, e^{-W[\phi, \phi_A]}\, \sum_n\, Z_n  \;,
\label{Z} \\
&& Z_n= \int [Dm]_n \, \exp \left[ {    - \sum_{k=1}^{n}\int_0^{L_{k}} ds_k \;
\big( \cdot  \big)_k   }\right]\     \makebox[.5in]{,}  u_\mu =\frac{dx_\mu}{ds}  \in S^3  \;, \nonumber \\
&& \big( \cdot \big)  = \tau + \frac{1}{2} (\bar{z}_c
\dot{z}_c - \dot{\bar{z}}_c z_c)+ \frac{1}{2\xi}\, \dot{u}_\mu
\dot{u}_\mu  - i\,  u_\mu\,  I_{A}\, \Lambda_{\mu}^{A}(x) +
\phi(x) +I_{A}\, \phi^{A}(x)  \;,    \nonumber
\end{eqnarray}
\begin{eqnarray}
[Dm]_n &\equiv& \frac{1}{n!}\int_{0}^{\infty}\;
\frac{dL_{1}}{L_{1}}\frac{dL_{2}}{L_{2}}...\frac{dL_{n}}{L_{n}} \;
\int\; dv_1 dv_2 \dots dv_n  \int  [Dv(s_1)_{v_1,v_1}^{L_1}   \dots [Dv(s_n)]_{v_n,v_n}^{L_n} \nonumber
\end{eqnarray}
\[
v: x,u, z \makebox[.5in]{,}  
 dv = d^{4}x \, d^3u\, dz\, d\bar{z}  \;,
\]
where  $n$ sums over the number of loops and
$W$ encodes some correlations among them; in particular, excluded volume effects (density-density interactions) are implemented with a $\phi^2$-term. Similarly,  
(magnetic) color-dependent density interactions are introduced by means of a $\phi_A ^2$-term in W. The measure $ [Dv(s)]_{v,\, v}^{L} $ refers to a single (smooth) loop of size $L$ that starts and ends at a given set of variables $v=x,u,z$\,; it  integrates over every possible shape. The associated weight $q(v,v,L)$ can be obtained from the end-to-end probability  for an open line $q(v,v_0,L)$, after identifying the initial and final points. In this manner, the sum over loops becomes,


\begin{equation}
\sum_n Z_n =   e^{  \int_{0}^{\infty}\frac{dL}{L}\;  \int  dv \,  q(v,v,L)  }
   \makebox[.3in]{,}
q(v,v_0,L) =  \int  [Dv(s)]_{v,v_0}^L \, e^{-  \int_0^{L} ds \,   
( \cdot  )} \;.
\label{endtend}
\end{equation}

\begin{figure}[h!]
\hspace{2.1cm}
\centering
\includegraphics[scale=.15, bb = -83 0 600 850]{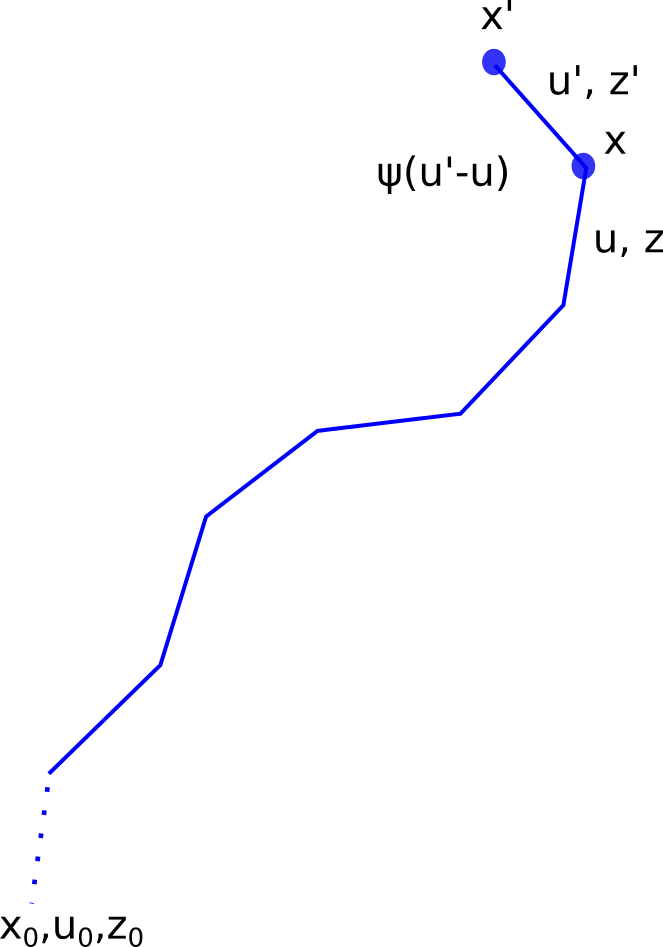}
\caption{Polymer growth, from $M$ to $M+1$ monomers.}
\label{fCHK}       
\end{figure}

The effective field model was obtained as follows. 
The probability $q(v,v_0,L) $ to start at $x_0$, with tangent $u_0$ and $z_0$, and end at $x$ with $u$, $z$ can be studied  by relying on the equilibrium theory of inhomogeneous polymers \cite{ref24}.  For this aim, we considered a  Chapman-Kolmogorov recurrence relation for diffusion in $v$-space (polymer growth),
\begin{eqnarray}
\label{recesi}
&& q_{j}(x,x_0,u,u_0, \bar{z},z_0) = \int d^{4}x' d^{3}u'\, dz'd \bar{z}' \;
e^{-\tau\Delta L}\, e^{(\bar{z}-\bar{z}')\cdot z'} \,  \times
\nonumber \\
&& ~~~~
\psi(u-u') \, e^{-\omega(x,u, \bar{z},z')\Delta L} \, \delta(x - x' -
u\Delta L)\, q_{j-1}(x',x_{0},u',u_{0}, \bar{z}',z_0)  \/\;, \nonumber \\
&& \psi(u-u')=\mathcal{N}\, e^{-\frac{1}{2\xi}\Delta L\left(\frac{u-u'}{\Delta
L}\right)^{2}} \makebox[.3in]{,}  \omega(x,u, \bar{z},z')= \phi(x) -i\,  u_\mu \Lambda_{\mu}^{A}(x) T^{A}_{cd}\,
\bar{z}^{c} {z'}^{d} +
\phi^{A}(x) T^{A}_{cd}\, \bar{z}^{c} {z'}^{d}  \;.  \nonumber
\end{eqnarray}
Taking the initial condition,
$ q_{0}(x,x_{0},u, u_{0}, \bar{z},z_0)=\delta(x-x_{0})\, \delta(u-u_{0})\,
e^{\bar{z}\cdot z_{0}} $, 
a  discretized version of $q$ for a size $L = M\,  \Delta L$ polymer
was obtained upon $M$ iterations.
Then, using the recurrence, a Fokker-Plank equation was derived  by
relating the probability to get $x$, $u$, $z$ with $M$ monomers
and the probability to get $x'= x+ u'\,  \Delta  L$,  $u'$, $z'$ with $M+1$ (see fig. \ref{fCHK}),
\begin{equation}
\partial_{L}q = \left[-\tau-\phi(x) + \frac{\xi}{\pi}\, \hat{L}^{2}_{u}-u_{\mu}\partial_{\mu}+ (i\, 
u_{\mu}\Lambda_{\mu}^{A}-
\phi^{A})\,M^{A}_{cd}\, \bar{z}^{c}\frac{\partial}{\partial \bar{z}^{d}} \right]
q \;.
\label{FPtot}
\end{equation}
 
\vspace{-1cm}

The path-integral over the $z$-sector  (cf.  eq.  (\ref{endtend})) can also be written as a transition between coherent color states, 
\begin{eqnarray} 
q(v,v_0,L) =  \int [Dx(s)]_{x,x_0}^{L} [Du(s)]_{u,u_0}^{L} \, \langle z| P \left\{ e^{-\int_0^L ds\, \hat{H}(s)}  \right\} |z_{0}\rangle \;,
\label{qeq}  \\
\hat{H}(s)=\Big(   \tau +  \frac{1}{2\xi}\, \dot{u}_\mu \dot{u}_\mu  + \phi(x(s)) \Big)\, \hat{I} -i\, u_\mu \Lambda_{\mu}^{A}(x(s)) \, M^{A}_{cd}\,
\hat{a}^{\dagger}_{c}\hat{a}_{d} + \phi^{A}(x(s))\, M^{A}_{cd}\,
\hat{a}^{\dagger}_{c}\hat{a}_{d} \;,
\label{Heq}
\end{eqnarray}
where $|z\rangle $ stands for the overcomplete basis of coherent states $|z_1, \dots, z_D \rangle$, $\hat{a}_c  |z\rangle = z_c | z \rangle $, in a linear space of general (magnetic) color states \cite{OST}. Therefore, 
\begin{equation}
\sum_n Z_n =  \exp{  \int_{0}^{\infty}\frac{dL}{L}\;  \int  d^4x\, d^3u \, \int [Dx(s)]_{x,x}^{L} [Du(s)]_{u,u}^{L} \,  {\rm Tr}\, P \left\{ e^{-\int_0^L ds\, \hat{H}(s)}  \right\} } \;.
\end{equation}
This trace is originated from the $dz \, d\bar{z}$ integration. Switching to the occupation number basis $|N\rangle$, \footnote{$N$ denotes the tuple of occupation numbers $(N_1, \dots, N_D)$.} 
\begin{equation}
 \sum_n Z_n  =  \exp{  \int_{0}^{\infty}\frac{dL}{L}\;  \int  d^4x\, d^3u\,   \sum_{N} Q_{NN}} \;,
\label{sumZ}  
\end{equation}
\begin{eqnarray}
 ~~~~~~  ~~~~~ Q_{NM} &=&  \int [Dx(s)]_{x,x}^{L} [Du(s)]_{u,u}^{L} \, \langle N | P \left\{ e^{-\int_0^L ds\, \hat{H}(s)}  \right\}|M\rangle
 \nonumber \\ 
 & = & \int dz d\bar{z}\, dz_0 d\bar{z}_0  \,  e^{-   \frac{\bar{z}\cdot z}{2}} 
 e^{- \frac{\bar{z}_0\cdot z_0}{2}}\, 
\bar{\psi}_N(z)\, \psi_M(\bar{z}_{0}) \,   q(v,v_0,L) \;.
\end{eqnarray} 
A partial contribution to $\sum_N Q_{NN}$ comes from states $|N\rangle$ of the form $|0,\dots, 1,\dots,0\rangle$, with one nontrivial entry $N_a=1$, $a= 1, \dots, D$. They have well-defined color $a$, and are represented by the wavefunction $\bar{\psi} (z)= z^a$. Organizing the associated matrix elements in a reduced $D\times D$ matrix $Q|^{cd} =Q^{cd}$, 
eq. (\ref{FPtot}) was projected into  the reduced Fokker-Plank equation, 
\begin{equation}
\left[(\partial_L - (\xi /\pi)\, \hat{L}^{2}_{u}+(\tau +\phi )\,
1 + \phi^A  M_A  + u\cdot D\right] Q(x,x_0,u, u_{0},L)=0 \;.
\label{reduFP}
\end{equation}
\[
Q(x,x_0,u,u_0,0)= \delta(x-x_{0})\, \delta(u-u_{0})\, 1 \makebox[.5in]{,}
D_\mu = 1\, \partial_{\mu} - i\, \Lambda_{\mu}^{A}  M^{A} \;.
\]
In this manner, taking the semiflexible limit (small stiffness), we arrived at,
\begin{eqnarray} 
\lefteqn{  \int d^4x\, d^3u\, Q(x,x, u,u,L) \approx \int d^4x\,  \langle x| e^{-LO} | x \rangle
\makebox[.5in]{,}
O = - \frac{\pi}{12 \xi}\, D_\mu D_\mu + (\phi +\tau ) \, 1 + \phi^A  M_A  \;,  } 
\nonumber \\
&&   \sum_n Z_n =   \exp {\int_{0}^{\infty}\frac{dL}{L}\; \int d^{4}x \, d^3u \,
\sum_{a=1}^D Q_{aa}(x,x,u,u,L) + \dots} = \exp\, [{-{\rm Tr} \ln O}] \, \dots \nonumber \\ 
\label{Qeq}
\end{eqnarray}
The dots represent the contribution originated from other sectors, which produce effective fields carrying product representations of the original $D$-dimensional representation. Finally, performing the path-integral in eq. (\ref{Z}), and using the adjoint representation,  the sum over the ensemble was obtained in terms of an effective complex adjoint Higgs field  
$\zeta = \psi_1 + i\, \psi_2~$,~ 
 $
 Z = \int [D\psi]\, e^{-\int d^4x\, {\cal L}_{\rm eff} }  \;,
$
\begin{equation}
 {\cal L}_{\rm eff} =   \frac{1}{2} \langle D_\mu \psi_I ,
D^\mu \psi_I\rangle    + \frac{\mu^2}{2} \langle \psi_I,\psi_I \rangle  + \frac{\lambda}{4}\, \langle
\psi_I,\psi_I \rangle \langle \psi_J,\psi_J \rangle + \frac{\eta}{4}\, \langle
\psi_I\wedge \psi_J,\psi_I \wedge \psi_J\rangle  \;,
\label{Leq}
\end{equation}
where $\mu^2 \propto \tau \, \xi$,  and $\psi_I$, $I=1,2$ is a pair of Hermitian adjoint Higgs fields.

\subsection{Non Abelian coupling and group coherent states}

In the non-Abelian coupling (\ref{Balac}),  instead of using linear variables $z_c$  we can consider nonlinear (Gilmore-Perelemov) group coherent states for the adjoint representation,  
\[
| z\rangle   = \tilde{R}\, | u_\alpha \rangle  \makebox[.5in]{,}  \tilde{R} \in {\rm Ad} ({\rm SU}(N))\;,
\]
where $ | u_\alpha \rangle  $ is the weight vector for the highest weight $\vec{\alpha}$ of the adjoint representation. This is an overcomplete basis 
over a $D$-dimensional vector space ($D=N^2-1$),
$\int d(G/H) \, |z \rangle \langle z| = I  $. 
In this case, 
 \begin{equation}
    \frac{dx_\mu}{ds} \, I_A   \,   \Lambda_\mu^A  -i \, z_c \dot{\bar{z}}_c 
 = 
  \frac{dx_\mu}{ds} \,     {\rm Tr}\,  (\tilde{R}^{-1} \Lambda_\mu\, \tilde{R} 
+i\, \tilde{R}^{-1 }  \partial_\mu  \tilde{R} ) \, (\vec{\alpha}\cdot \vec{M})  \;,
 \end{equation}
and  instead of eqs. (\ref{qeq}) and (\ref{Heq}), 
we have \cite{Ox},
\begin{eqnarray}
&& q(v,v_0,L) =  \int [Dx(s)]_{x,x_0}^{L} [Du(s)]_{u,u_0}^{L} \,  \langle z|  P \left\{ e^{-\int_0^L ds\, H(s)}   \right\} |z_{0}\rangle   \;,  \\
&&~~~ ~~~ ~ ~ H(s)=\Big(   \tau +  \frac{1}{2\xi}\, \dot{u}_\mu \dot{u}_\mu  + \phi(x(s)) \Big)\, I -i\, u_\mu \Lambda_{\mu}^{A}(x(s)) \, M_{A} + \phi^{A}(x(s))\, M_{A} \;.  \nonumber
\end{eqnarray}
In this respect, note that the path-integral over the $z$-sector in  eq.  (\ref{endtend}) (when $\phi_A=0$) is proportional to  the Petrov-Diakonov representation of transition elements of a (dual) Wilson line, computed between group coherent  states. Then, eq. (\ref{Qeq}) gets replaced by, 
\begin{eqnarray}
&& \sum_n Z_n =   \exp {\int_{0}^{\infty}\frac{dL}{L}\; \int d^{4}x \, d^3u \,
\sum_{a=1}^D \Gamma_{aa}(x,x,u,u,L)} \;,  \\
&& ~~~~~ ~~~  \Gamma_{ba}(x,x_0,u,u_0,L) =  \int [Dx(s)]_{x,x_0}^{L} [Du(s)]_{u,u_0}^{L} \,  
P \left. \left\{ e^{-\int_0^L ds\, H(s)}   \right\}\right|_{ba} 
\;.
\label{Gmat}
\end{eqnarray}
The  discretized version of  this path-integral, in matrix form,  
can again be obtained from  a Chapman-Kolmogorov recurrence relation. Moreover, using again the recurrence to relate $M$ and $M+1$ steps, we can show that $\Gamma$ satisfies  eq. (\ref{reduFP}), with $\Gamma$ in the place of $Q$,  see ref. \cite{Ox}. 
Then, the consideration of Gilmore-Perelemov group coherent states leads directly to the former reduced Fokker-Planck equation, and a representation only based on effective adjoint fields is obtained, without product representations.

\subsection{Higgs potentials and flavor symmetry}
 
Now, let us  discuss some specific features of ${\rm SU}(N) \to {\rm Z}(N)$
YMH models  motivated from the ensemble point of view. 
In the ${\rm SU}(2)$ case, the simplest model is based on a pair of real adjoint scalars \cite{deVega}-\cite{HV},  
\begin{eqnarray}
\lefteqn{V_{\rm Higgs}(\psi_1,\psi_2) = \frac{\mu_1^2}{2}\, \langle \psi_1, \psi_1\rangle + 
\frac{\mu_2^2}{2}\, \langle \psi_2, \psi_2\rangle  }  + \nonumber \\
&& + \frac{\lambda_1}{4} \langle \psi_1, \psi_1\rangle^2 + 
\frac{\lambda_2}{4}  \langle \psi_2, \psi_2\rangle^2  + \frac{\gamma}{2} \langle \psi_1, \psi_1\rangle \langle \psi_2, \psi_2\rangle +\frac{\beta}{2}\, \langle \psi_1, \psi_2 \rangle^2
\end{eqnarray}
($\lambda_1, \lambda_2 > 0$). To drive ${\rm SU}(2) \to {\rm Z}(2)$ SSB , the following conditions must be satisfied: 
$
\mu_1^2 , \mu_2^2 < 0 $, $
\lambda_2\, \frac{\mu_1^2}{\mu_2^2} > \gamma + \beta $, 
$
\lambda_1\, \frac{\mu_2^2}{\mu_1^2} > \gamma + \beta 
$. Here, the last term is essential for ${\rm SU}(2) \to {\rm Z}(2)$ SSB, as for nontrivial $\langle \psi_1, \psi_1 \rangle$, $\langle \psi_2, \psi_2 \rangle$ it favors a minimization with linearly independent fields, $\psi_1$ and $\psi_2$. Then, in this case, the  SSB phase cannot display an additional  
global ${\rm U}(1)$ flavor symmetry, 
\begin{equation}
\psi_1' = \cos \omega \; \psi_1 + \sin \omega \; \psi_2 
\makebox[.5in]{,}
\psi_2' = -\sin \omega \; \psi_1 + \cos \omega \; \psi_2  \;.
\end{equation}
This symmetry would be present for  $\mu_1^2=\mu_2^2 $, $\lambda_1 =\lambda_2 =\gamma$ and, necessarily, vanishing $\beta $. On the other hand, any effective field model derived from a loop ensemble is expected to posses this symmetry, which is related to the equivalence  between loops with different orientations. In particular, the potential in eq. (\ref{Leq}) is flavor symmetric,
\begin{equation}
V_{\rm Higgs}(\psi_1,\psi_2) =    \frac{\mu^2}{2} \left[ \langle \psi_1,\psi_1 \rangle + 
\langle \psi_2 ,\psi_2 \rangle \right]  + \frac{\lambda}{4}\, \left[ \langle \psi_1,\psi_1 \rangle + 
\langle \psi_2 ,\psi_2 \rangle \right]^2 + \frac{\eta}{2}\, \langle
\psi_1\wedge \psi_2,\psi_1 \wedge \psi_2\rangle  \;.
\end{equation} 
If $\mu^2 < 0$, $\lambda , \eta > 0$, there is no ${\rm SU}(2) \to {\rm Z}(2)$ SSB, as the last term tends to align $\psi_1 $ and $\psi_2$. However, this phase does appear for $\eta < 0$,  $2\lambda + \eta > 0$. 
Another possibility is to embed ${\rm U}(1)$ in ${\rm SO}(3)= {\rm Ad}({\rm SU}(2))$ 
by including a third adjoint field $\psi_3\,$. In  general, ${\rm Ad} ({\rm SU}(N))$ flavor symmetry was implemented in ref. \cite{conf-qg}. For this objective, we considered $N^2-1$ flavors, $ I \to A=1,\dots, N^2-1$ and the Higgs potential,
\begin{equation}
V_{\rm Higgs} = c+ \frac{\mu^2}{2}\, \langle \psi_A ,\psi_A \rangle   +\frac{\kappa}{3} \,f_{ABC}\, \langle \psi_A \wedge \psi_B,\psi_C \rangle
+ \frac{\lambda}{4}\, \langle \psi_A \wedge \psi_B,\psi_A\wedge \psi_B \rangle \;.
\label{nonAb}
\end{equation}
When $\mu^2 < \frac{2}{9}\frac{\kappa^{2}}{\lambda}$, the absolute minima are given by Lie bases associated with  structure constants 
$f_{ABC}$, thus driving ${\rm SU}(N) \to {\rm Z}(N)$ SSB. These vacua have a color-flavor locking symmetry ${\rm Ad}({\rm SU} (N))_{\rm C+F}$ similar to that proposed in refs. \cite{David}, \cite{it},  \cite{GSY}. In addition, at $\mu^2=0$  center string field equations get Abelianized  \cite{DL}, with $N-1$ fields pointing along the Cartan directions, and frozen at vacuum values.  
The remaining $N(N-1)$ real adjoint fields are along off-diagonal directions. They can be combined as $\phi_{\alpha}\, E_\alpha $, where 
$E_\alpha$ are root vectors (with positive weight $\vec{\alpha}$) and 
 $\phi_{\alpha_1}$, $\phi_{\alpha_2}, \dots$ are $N(N-1)/2$ complex scalar fields. This Abelianized content makes contact with effective models for monopole ensembles in pure YM,  based on Abelian projection and Abelian dominance \cite{antonov}.

\section{Conclusions}

In this work, we initially discussed some physical properties of effective ${\rm SU}(N) \to {\rm Z}(N)$ YMH models  emphasizing the important role of valence gluons, which are represented by confined (dual) monopoles. Among them, we list: i) The possibility of hybrid mesons, currently searched by the GlueX collaboration, ii) The correct picture for the adjoint string breaking between adjoint quarks, iii)  
Difference-in-areas for doubled pairs of fundamental ${\rm SU}(2)$ quarks, understood as the minimum energy for doubled quark-antiquark pairs. 
 
Next, we reviewed how adjoint fields are naturally generated as an effective description of ``adjoint'' loops  in $4D$, with tension, stiffness, and (magnetic) color degrees of freedom.  On the other hand, monopole loops with adjoint charges have been detected in center vortex ensembles that capture the path-integral measure in infrared YM theories. This gives further support to look for dual superconductor models with adjoint Higgs fields. In this respect, we discussed the simplest ensemble of loops in  ${\rm SU}(2)$ showing that the symmetry between different loop orientations is manifested as a ${\rm U}(1)$ flavor symmetry,  which can   guide the construction of the effective model. This could also be accommodated in a larger ${\rm Ad}({\rm SU}(2))$ flavor symmetry realized in recently studied models, and extended to the case of ${\rm SU}(N)$. 

Answering what is the appropriate field content and symmetries will require further steps to characterize ensembles  in pure YM, elucidating the origin of magnetic non Abelian d.o.f. and the role of center vortices in the model construction. The latter is a particularly difficult task as center vortices are two-dimensional objects (for a treatment in the continuum, see refs. \cite{ref11}-\cite{EQR}). On the other hand, this is the component  that can pierce Wilson loops to produce ($N$-ality dependent) center elements.
Here, we anticipated some simpler steps. Namely, the  projection of  effective fields into the adjoint representation, without  product representations, by coupling magnetic color degrees via Gilmore-Perelemov group coherent states. Indeed, in ref. \cite{Ox}, we will show that these degrees are naturally originated by combining center gauges  that detect magnetic defects in the continuum  \cite{OS-det} together with a non Abelian Hodge decomposition.

\subsection*{Acknowledgements}

The Conselho Nacional de Desenvolvimento Cient\' {\i}fico e Tecnol\'ogico (CNPq) is acknowledged for the financial spport.


\begin{thebibliography}{bibi}

\bibitem{N} Nambu, Phys. Rev. {\bf D10}, (1974) 4262.

\bibitem{M} Mandelstam, Phys. Rep. {\bf 23C} (1976) 245. 

\bibitem{3} G. 't Hooft, Nucl. Phys.  B138 (1978) 1.


\bibitem{Su-90} S. Maedan, Y. Matsubara and T. Suzuki, Prog. of Theor. Phys. {\bf 84} (1990) 130.

\bibitem{su3-Suzuki} Y. Koma, E. M. Ilgenfritz, H. Toki, and T. Suzuki, Phys. Rev. {\bf D64} (2001) 011501(R). 

\bibitem{Ingel} Y. Koma, M. Koma, E. M. Ilgenfritz and T. Suzuki, Phys. Rev. {\bf D68} (2003) 114504.  
 
\bibitem{Baker-90}  M. Baker, J. S. Ball and F. Zachariasen, Phys. Rev.
\textbf{D41} (1990) 2612.

\bibitem{Baker-91} M. Baker, J. S. Ball and F. Zachariasen, Phys. Rev.
\textbf{D44} (1991) 3328.

\bibitem{Baker-Dosch}  M. Baker, N. Brambilla, H. G. Dosch, and A. Vairo, Phys. Rev.
\textbf{D58} (1998) 034010. 

\bibitem{David} A. Hanany and D. Tong, JHEP {\bf 0307} (2003) 037

\bibitem{HDavid} A. Hanany and D. Tong, JHEP 0404 (2004) 066.

\bibitem{Dav} D. Tong, Phys. Rev. {\bf D69} (2004) 065003. 
 
\bibitem{vortices-d} D. Tong, Ann. of Phys. \textbf{324} (2009) 30. 

\bibitem{it} R. Auzzi, S. Bolognesi, J. Evslin, K. Konishi and A. Yung, Nucl. Phys. B 673 (2003) 187.

\bibitem{notes} K. Konishi, Lect. Notes Phys. {\bf 737} (2008) 471.

\bibitem{GSY} A. Gorsky, M. Shifman and A. Yung, Phys. Rev. {\bf D71} (2005) 045010. 

\bibitem{AS} A. Armoni and M. Shifman, Nucl. Phys. {\bf B671} (2003) 67.

\bibitem{Nonajun} M. Shifman and A. Yung, Phys. Rev. D 70, 045004 (2004).

\bibitem{Konishi-Spanu} K. Konishi and L. Spanu, Int. J. Mod. Phys. {\bf A18} (2003) 249. 

\bibitem{Marco} Marco A. C. Kneipp and Patrick Brockill, Phys. Rev. {\bf D64} (2001) 125012.

\bibitem{Mgluon} Marco A. C. Kneipp, Phys. Rev. {\bf D76} (2007) 125010. 

\bibitem{conf-qg} L. E. Oxman, J. High Energy Phys. \textbf{03} (2013) 038.

\bibitem{ref19}

G. Mack and V. B. Petkova,  Ann. Phys. \textbf{123} (1979) 442; \textit{ibid.} \textbf{125} (1980) 117.

\bibitem{ref14}

L. Del Debbio, M. Faber, J. Greensite and S. Olejnik,  Phys. Rev. \textbf{D55} (1997) 2298.

\bibitem{ref15}

K. Langfeld, H. Reinhardt and O. Tennert, Phys. Lett. \textbf{B419} (1998) 317; \textit{ibid.} \textbf{B431} (1998) 141.

\bibitem{ref11}

M. Engelhardt and H. Reinhardt,  Nucl. Phys. \textbf{B567} (2000) 249.

\bibitem{reinhardt} H. Reinhardt, Nucl. Phys. {\bf B628}
(2002) 133.
 
\bibitem{EQR} M. Engelhardt, M. Quandt, and H. Reinhardt, Nucl. Phys. {\bf B685} (2004) 227.

\bibitem{ref17}

Ph. de Forcrand and M. Pepe, Nucl. Phys. \textbf{B598} (2001) 557.

\bibitem{ref18}

F. V. Gubarev, A. V. Kovalenko, M. I. Polikarpov, S. N. Syritsyna and V. I. Zakharov, Phys. Lett. \textbf{B574} (2003) 136.


\bibitem{greensite} J. Greensite, Prog. Part. Nucl. Phys. {\bf 51} (2003) 1.

\bibitem{book-G} J. Greensite, {\it An introduction to the confinement problem} (Springer, Berlin-Heidelberg, 2011).

\bibitem{Biagio} B. Lucini and M. Teper, Phys. Lett. {\bf B501}  (2001) 128.

\bibitem{deVega} H. J. de Vega, Phys. Rev. \textbf{D18} (1978) 2932.

\bibitem{Fidel2} H. J. de Vega and F. A. Schaposnik, Phys. Rev. {\bf D34} (1986) 3206.
 
\bibitem{HV} J. Heo and T. Vachaspati, Phys. Rev. \textbf{D58} (1998) 
065011. 

\bibitem{OST} L. E. Oxman, G. C. Santos Rosa and B. F. I. Teixeira,  J. Phys. A: Math. Theor. {\bf 47} (2014) 305401.

\bibitem{HK} M. Hindmarsh and T. W. B. Kibble, Phys. Rev. Lett. 55 (1985) 2398.

\bibitem{L2008} L. E. Oxman, JHEP {\bf 0812} 2008 089. 

\bibitem{gluex} GlueX website: http://www.gluex.org/GlueX/Home.html

\bibitem{DL} L. E. Oxman and D. Vercauteren, arXiv:1603.01105.

\bibitem{dFK} S. Kratochvila and  P. de Forcrand, Nucl. Phys. {\bf B671} (2003) 103.

\bibitem{GR} Jeff Greensite and Roman Höllwieser, Phys. Rev. {\bf D91} (2015) 054509.

\bibitem{bar-sam} K. Bardakci and S. Samuel, Phys. Rev. {\bf D18} (1978) 2849.

\bibitem{HSi} M. B. Halpern and W. Siegel, Phys. Rev. {\bf D16 } (1977) 2486. 

\bibitem{ref26} A. P. Balachandran, P. Salomonson, B. Skagerstam and J. Winnberg, 
 Phys. Rev.  \textbf{D15} (1977) 2308.
 
\bibitem{ref24} G. H. Fredrickson,  \textit{The Equilibrium Theory of Inhomogeneous Polymers} ($2^{nd}$edition, Clarendon Press, Oxford, 2006), p. 452. 
 
\bibitem{antonov} D. Antonov, Surveys High Energ. Phys. {\bf 14} (2000) 265.

\bibitem{Ox} L. E. Oxman, in preparation.

\bibitem{OS-det} L. E. Oxman and G. C. Santos-Rosa,
Phys. Rev. {\bf D92} (2015) 125025.


\end{thebibliography}
\end{document}